\documentclass[twocolumn,showpacs,preprintnumbers,amsmath,amssymb,superscriptaddress]{revtex4}

\usepackage{graphicx}
\usepackage{dcolumn}
\usepackage{bm}

\begin{document}


\title{Temperature-dependent transition from injection-limited to space-charge-limited current in metal-organic diodes}

\author{Yi Zheng}
  \email{phyzy@nus.edu.sg}
 \author{Andrew T. S. Wee}
  \affiliation{Department of Physics, National University of
Singapore, 2 Science Drive 3, Singapore 117542}
\author{Cedric Troadec}
\author{N. Chandrasekhar}
  \email{n-chandra@imre.a-star.edu.sg}
\affiliation{Institute of Material Research and Engineering (IMRE),
3 Research Link, Singapore 117602}

\date{\today}

\begin{abstract}
Based on the assumption that the contact barrier height determines the current flow in organic semiconductor-based electronic devices, charge injection at metal-organic (MO) interfaces has been extensively investigated, while space-charge conduction in organic bulk is generally overlooked. Recent theoretical modeling and simulation have pointed out that such a simplification is questionable due to the hopping nature of charge injection and hopping-related space-charge conduction. Here we show experimentally that charge transport in metal-organic diodes is complex interplay between injection-limited current (ILC) and space-charge-limited current (SCLC). We report the experimental observation of ILC-to-SCLC transition in Ag/pentacene/Ag diodes as a function of temperature.
\end{abstract}

\pacs{72.80.Le,73.40.Sx,72.20.-i,72.60.+g}
\maketitle
The operations of all organic electronic devices, such as organic light-emitting diodes (OLEDs) and organic thin-film transistors (OTFTs), depend on controlling current flow in these devices. Two competing electronic processes, namely charge injection from the contacts and space charge conduction in the organic bulk, should be taken into account when analyzing the charge transport measurements. For a conventional treatment, the contact barrier height (BH) is the only parameter determining which electronic process is the limiting factor for current flow. For a Schottky-type contact with BH higher than 0.25-0.3 eV \cite{Pope99Organic}, the current is believed to be controlled by interfacial charge injection, which can be modeled by the thermionic emission model \cite{Bethe42MIT} or the Schottky diffusion theory \cite{Schottky31PhyZ}. With lower barriers or nearly ohmic contacts, an intrinsic SCLC \cite{Lambert70CurrInjSolids} is expected with $J_{SCL}=(\frac{9}{8})\epsilon \epsilon_{r}\mu \frac{V^{2}}{L^{3}}$.

\begin{figure}
\begin{center}
\includegraphics[width=3.3in]{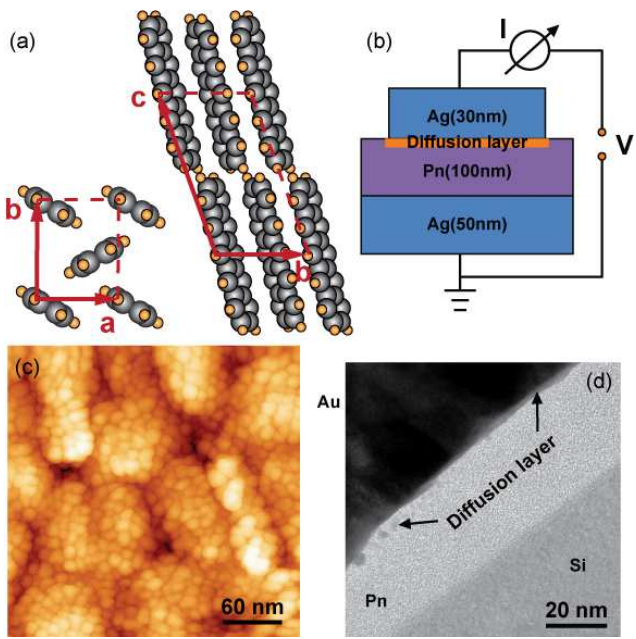}
\end{center}
\caption{(a) Typical crystal structure of planar $\pi$-conjugated semiconductors. Two inequivalent molecules are herringbone-packed in the $a$-$b$ plane with relatively strong intermolecular $\pi$-$\pi$ interaction, while along the $c$-axis molecules are bound by van der Waals force with negligible electronic coupling. (b) Sample geometry of Ag/Pn/Ag diodes and the experimental setup for charge transport measurements. (c) Scanning tunneling microscopy shows that the top Ag forms a continuous film with a typical granular surface. (d) Metal diffusion layer at MO interface due to direct metal evaporation on organic thin film. TEM cross-section of Au/Pn/Si. \label{Fig01}}
\end{figure}

Such a treatment is based on high charge carrier mobility in the semiconductor and an exponential dependence of charge injection on the contact BH. Both are not valid for organic semiconductors. Among thermally evaporated planar $\pi$-conjugated molecules, purified single crystals of pentacene (Pn) have the highest room-temperature (RT) mobility of $\sim35\, \mathrm{cm^{-2}V^{-1}s^{-1}}$ \cite{Jurchescu04APL}, which is rather low compared to the common high-mobility Si and GaAs. For device applications, organic thin films are more preferred, in which mobility is $\sim1\,\mathrm{cm^{-2}V^{-1}s^{-1}}$ or even lower. The mobility in organic is also highly anisotropic, closely related to $\pi$-$\pi$ interactions induced by molecular packing. Fig. \ref{Fig01}a shows the typical crystal structure of small planar molecules. In the $a$-$b$ plane, band transport \cite{Koch06PRL} and medium mobility (0.1 to 35 $\mathrm{cm^{-2}V^{-1}s^{-1}}$) can be found due to sufficient $\pi$ orbital overlapping. In contrast, along the $c$-axis, hopping-type transport and very low mobility (comparable to disordered polymer semiconductors, $\sim10^{-6}\, \mathrm{cm^{-2}V^{-1}s^{-1}}$) are expected, since $\pi$-$\pi$ interaction along this direction is negligible.

Besides the low mobility, it is generally found that current injection at MO interfaces is over estimated when using the classical thermionic or diffusion theory. The charge injection at MO interfaces is also not very sensitive to temperature changes, in contrast to the strong temperature-dependent thermionic/diffusion injection. Based on these experimental findings, it has been proposed that current injection at MO interfaces is by hopping instead of thermionic/diffusion process \cite{Abkowitz95APL,Arkhipov98JAP}. In Arkhipov \textit{et al}'s treatment, hopping injection at MO interfaces is a two-step process \cite{Arkhipov98JAP}. Charge carriers first jump from the Fermi surface ($E_\mathrm{F}$) of metal to localized states in organic near the interface. The injected charge carriers will either flow back to metal or keep on hopping forward with the assistance of an external electric field. In this model, the total hopping injection current is
\begin{equation}\label{HoppingInjction}
J=q\nu \int_{a}^{\infty}dx_{0}\exp(2\gamma x_{0})w_\mathrm{esc}\int_{-\infty}^{\infty}dE \,\text{Bol}(E)g[E-U(x_{0})],
\end{equation}
where $w_\mathrm{esc}$ and $g[E]$ are the one-dimensional Onsager escape probability and density-of-states function respectively, and $\text{Bol}(E)=\left\{ \begin{array}{ll}
\exp(-\frac{E}{kT}), \, E>0,\\
1, \qquad \qquad E<0
\end{array} \right.$ \cite{Arkhipov98JAP}.

Unlike the conventional treatment, Eq. \eqref{HoppingInjction} only defines the injection boundary conditions by $J(0)=J_{hop}(F_{0})$ and $F(0)=F_{0}$. The current-voltage ($I$-$V$) characteristics of metal-organic diodes should be calculated by solving equations \cite{Lambert70CurrInjSolids}
\begin{equation}\label{Hopping-SCLC}
   J(x)=q\mu_{p}p(x)F(x) \, \mathrm{and} \, \frac{\varepsilon_{0}\varepsilon_{r}}{q}F'(x)=p(x),
\end{equation}
with the hopping injection boundary conditions.

In general, the hopping model gives better description of charge injection at MO interfaces than the classical models \cite{Barch99PRB,Vissenberg01APL}, and is confirmed by Monte Carlo simulation \cite{Arkhipov99PRB59}. More interestingly, it predicts complex interplay between ILC and SCLC as a function of temperature and electric field. Because the hopping injection has a much weaker temperature dependence, current in MO diodes could be gradually controlled by space-charge conduction even with a high injection BH of 1 eV as temperature drops \cite{Arkhipov03APL83}. Such interplay and transition between SCLC and ILC have not yet been observed by experiments, which requires well-controlled MO interfaces and temperature-dependent charge transport measurements.

\begin{figure}
\begin{center}
\includegraphics[width=3.0in]{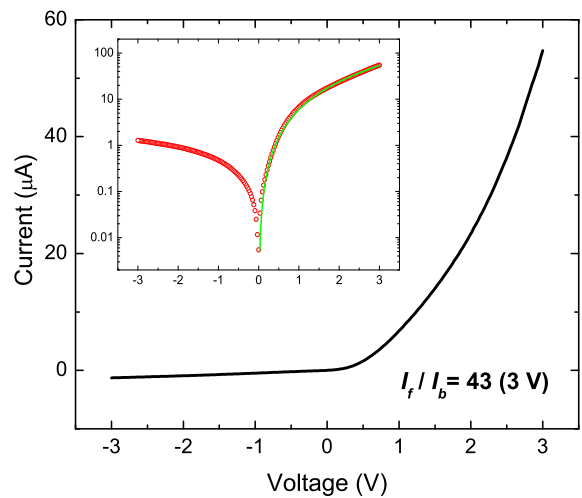}
\end{center}
\caption{RT $I$-$V$ characteristics of one Ag/Pn/Ag diode. The inset shows the same curve in log scale. The polarity of the bias is defined in Fig. \ref{Fig01}b.\label{Fig02}}
\end{figure}

The samples used in this study is prepared into a sandwich structure of Ag/Pn/Ag (Fig. \ref{Fig01}b). All three layers are prepared by direct thermal evaporation in a time sequence of Ag, Pn and Ag. Previous study has found that Pn molecules stand up on Ag surfaces and form an ordered ``thin-film phase'' structure \cite{Zheng07Langmuir}. The resulting Pn-on-Ag interfaces have a hole injection barrier of $\sim0.6$ eV \cite{Zheng07Langmuir}. The direct evaporation of Ag on Pn thin films (Fig. \ref{Fig01}c) creates heavily diffuse and nearly ohmic Ag/Pn interfaces. This can be seen both from the transmission electron microscopy (TEM) cross-section (Fig. \ref{Fig01}d), and the $I$-$V$ characteristics shown in Fig. \ref{Fig02}, in which the forward current (hole injection from the top Ag, $I_\mathrm{f}$) can approximately be fitted by the SCLC model. The ratio between $I_\mathrm{f}$ and the backward current (hole injection from the bottom Ag, $I_\mathrm{b}$) is more than 40 at a bias of $\pm3$ volt. Thus, we have an ideal system with space-charge-limited $I_\mathrm{f}$ and injection-limited $I_\mathrm{b}$ at RT to test Arkhipov \textit{et al}'s theory.

\begin{figure*}
\begin{center}
\includegraphics[width=4.7in]{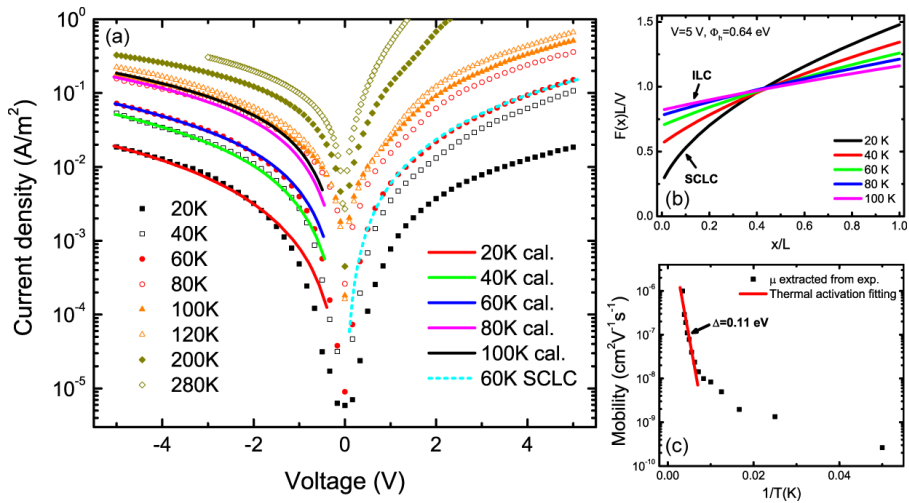}
\end{center}
\caption{(a) Temperature-dependent $I$-$V$ characteristics of the Ag/Pn/Ag diode. Different symbols are experimental $IV$s at different temperatures. Solid lines are the calculated $I_\mathrm{b}$. The dashed cyan line gives one example of the intrinsic SCLC fitting to $I_\mathrm{f}$. The $I_\mathrm{f}/I_\mathrm{b}$ ratio increases as a function of temperature due to SCLC-to-ILC transition in $I_\mathrm{b}$. (b) Spatial distribution of the electric field in the Ag/Pn/Ag diode as a function of temperature. A transition form ILC to SCLC can be clearly seen. (c) Hole mobility, extracted from the SCLC data fitting of $I_\mathrm{f}$, as a function of inverse temperature. The red line is a thermal-activation fitting of the mobility curve with $\Delta=0.11 \,\mathrm{eV}$ and $\mu_{0}=4.13\times10^{-5} \, \mathrm{cm^{2}V^{-1}s^{-1}}$. \label{Fig03}}
\end{figure*}

To do this, we first cooled down the samples to 5 K using liquid helium. The sample temperature was then adjusted to different setpoints using a Lakeshore 430 controller with temperature fluctuations less than 100 mK ($<10$ mK below 77 K). Conventional $IV$s were measured at each setpoint from 20 K to 280 K in a step of 20 K. A Keithley 6430 has been used for $I$-$V$ characterizations. All measurements were done in darkness and in a base vacuum level of $2\times10^{-5}$ mbar.

The temperature-dependent $I$-$V$ characteristics of Ag/Pn/Ag diodes are summarized in Fig. \ref{Fig03}a. At 20 K, $I_\mathrm{f}$ and $I_\mathrm{b}$ under the same bias is nearly identical. As temperature goes up, $I_\mathrm{f}$ increases much faster than $I_\mathrm{b}$, leading to increasing $I_\mathrm{f}/I_\mathrm{b}$ ratios from $\sim1$ at 20 K to $\sim15$ at 200 K. This observation agrees with the theoretical prediction that hopping injection has a much weaker temperature dependence than SCLC.  Above 200 K, the bias range is limited to $\pm\,3$ V to prevent the modification of O/M and M/O interfaces by high current density (only the curve at 280 K was shown for clarity).

We now focus on the transition occurs between 20 K and 100 K. We first fit $I_\mathrm{f}$ at each temperature setpoint using the intrinsic SCLC model, with one example for 60 K shown as the dashed cyan line in Fig. \ref{Fig03}a. The fitting process yielded the temperature-dependent hole mobility, which was further put into Eq. \eqref{Hopping-SCLC} to calculate the backward $I$-$V$ curves with the hopping injection boundary conditions of $J(0)=J_{hop}(F_{0})$ and $F(0)=F_{0}$ calculated by Eq. \eqref{HoppingInjction}. The parameters used in the hopping calculation are $\sigma=0.28\,\mathrm{eV}$, $a=0.5\,\mathrm{nm}$, $\gamma=\frac{1}{5\, \mathrm{nm}}$, $\epsilon_\mathrm{r}=3$ and $\Phi_\mathrm{h}=0.64\,\mathrm{eV}$. We can see that the hopping-injection model \eqref{HoppingInjction} and space-charge conduction equations \eqref{Hopping-SCLC} give quite a good description of the backward current as a function of voltage, except in the range below 1.5 V, where the current flow should be closely related to charge traps \cite{Pope99Organic}. Note that the $\Phi_\mathrm{h}$ used is slightly higher than the number obtained by UPS measurements, but consistent with the ballistic hole emission spectroscopy (BHES) results of intrinsic Ag/Pn interface with negligible M/O diffusion \cite{Zheng09PRL-Diffusion}. The difference may come from the fact that UPS was done at RT, while the other two measurements were acquired at much lower temperatures, which may modify the molecular DOS distribution.

One important information we can get from the above hopping-(space-charge conduction) calculation is the spatial electric field distribution in the diodes at different temperatures, which is a straightforward indication of SCLC or ILC. For ILC, the electric field should be nearly constant over the whole diode, while it has a strongly inhomogeneous distribution for SCLC. For normalization, the electric field, F(x), is multiplied by the diode's thickness, L, and divided by the applied voltage. In Fig. \ref{Fig03}b, we show the temperature-dependent field distribution functions under 5 V bias. The results clearly show that at 20 K, the charge transport is SCLC with a strongly inhomogeneous distribution of the electric field. As temperature increases, the inhomogeneity in the field becomes less. At 100 K, the electric field distribution is linear over the whole diode, indicating that the charge transport is now controlled by the interfacial injection. Here, the slope of the field distribution function represents the space charge effect in an injection-limited current.

Thus, our results on Ag/Pn/Ag give the direct experimental support for the predicted temperature-dependent ILC-to-SCLC transition in metal-organic diodes. However, the transition temperature (around 20 K) is lower than expected. To understand this, we also plot the extracted hole mobility from the SCLC fitting of $I_\mathrm{f}$ as a function of inverse temperature (1/T) in Fig. \ref{Fig03}c. We can see thermal activation behavior in the high temperature region, while it is approximately linear below 160 K. The small activation energy, $\Delta=0.11\,\mathrm{eV}$, and the deviation from the thermal-activation curve may be closely related to the band charge transport in the $a$-$b$ plane, in which hole mobility increases with decreasing temperature \cite{Jurchescu04APL,Koch06PRL,Sakamoto07PRL}. It would be interesting if such a 2D band structure could be incorporated into the hopping injection model, which simply treats organic thin film as disordered hopping sites.

In conclusion, using temperature-dependent charge transport study on Ag/Pn/Ag diodes, we have proved that current flow in metal-organic diodes is interplay between ILC and SCLC. The weaker temperature dependence of hopping injection compared to space-charge conduction leads to a transition from ILC to SCLC at low temperature. By taking the space-charge effects into account, the backward $I$-$V$ characteristics of Ag/Pn/Ag diodes can be modeled in a self-consistent way by solving the space-charge conduction equations with the boundary conditions defined by the hopping injection model \cite{Note01}. We further point out that the anisotropic band formation in organic should be incorprated into the hopping injection model instead of treating organic thin films as disordered hopping sites. The results obtained in this study can be applied to other metal-organic diodes using planar molecule semiconductors.

\newpage
\end{document}